%% file: paper275.tex
\documentclass[12pt,a4paper,dvips]{article}
\usepackage{a4p}
\usepackage{cite,mcite}
\usepackage{graphicx,pspicture}
\usepackage{rotating}
\usepackage{physics}
\usepackage{l3_title,ifthen,Lep}

\journalname{Phys. Lett. B}

\preprint{2003-055}

\date{August 27, 2003}

\newlength{\capindent}
\newlength{\capwidth}
\newlength{\figwidth}
\newcommand{\icaption}[2][!*!,!]{\hspace*{\capindent}%
  \begin{minipage}{\capwidth}
    \ifthenelse{\equal{#1}{!*!,!}}%
      {\caption{#2}}%
      {\caption[#1]{#2}}
  \end{minipage}}

\newcommand{\GG}{\ensuremath{\gamma ^* \gamma ^*}}
\newcommand{\Gg}{\ensuremath{\gamma \gamma}}
\newcommand{\pz}{\ensuremath{\pi^0}}
\newcommand{\ks}{\ensuremath{\rm K^0_S}}
\newcommand{\ppm}{\ensuremath{\pi^{\pm}}}

\newcommand{\Wgg}{\ensuremath{W_{\Gg}}}

\newcommand{\sqs}{\ensuremath{\sqrt{s}}}

\newcommand{\dpt} {\ensuremath{d\sigma / dp_t}}

\begin{document}
\bibliographystyle{l3style}
      
\begin{titlepage}
\title{Inclusive Jet Production \\ 
in Two-Photon Collisions at LEP}
\author{The L3 Collaboration}

\begin{abstract}
Inclusive jet production, $\ee \ra \ee$ jet  X,
is studied using 560  pb$^{-1}$  of data collected at LEP with the L3 detector  
at centre-of-mass energies between 189 and 209 \GeV. 
The inclusive differential cross section
is measured using a $k_t$ jet algorithm
as a function of the jet transverse momentum, \pt , in the
range $3<\pt<50$ \GeV {} for a pseudorapidity, $\eta$, in the range $-1<\eta<1$.
This cross section is well represented by a power law.
For high \pt , the measured cross section is significantly higher 
than the NLO QCD predictions, as already observed for inclusive
\pipm {} and \pz {} production. 
\end{abstract}

\submitted 

\end{titlepage}

\section{Introduction}

Two-photon collisions are the main source of hadron production 
in the high-energy regime of LEP 
 via the process ${\rm e}^{+} {\rm e}^{-} \rightarrow {\rm e}^{+} {\rm e}^{-}
 \gamma ^{*}  \gamma ^{*}  \rightarrow 
 {\rm e}^{+} {\rm e}^{-}  hadrons$.
Hadrons with  high transverse momentum 
are produced by the direct QED process $\GG \ra \qqbar$ 
or by QCD processes originating from the partonic content of the photon.
Next-to-leading order (NLO) QCD calculations are available \cite{frixione,frixione2}
for inclusive jet production in quasi-real two-photon interactions.

\par
The L3 Collaboration published results on inclusive \pz , \ks {} \cite{pz} and charged hadron 
 \cite{ch} production in quasi-real two-photon collisions. 
The inclusive \pz {} and \ppm {} differential cross sections,
measured as a function of transverse momentum,
exhibit a clear excess over NLO QCD calculations \cite{kniehl} for large transverse momentum.
In this Letter, inclusive jet production is studied, in similar two-photon interactions, for a  
centre-of-mass energy  of the two interacting photons, \Wgg ,  
greater than 5 \GeV. 
The jets are measured 
in the transverse momentum range \mbox{$3 < \pt < 50 \GeV$} and in 
the pseudo-rapidity interval $|\eta| < 1$.
The analysis of jet production allows a comparison of the measurements 
to NLO QCD predictions, expected to be largely independent 
of fragmentation functions and hadronisation models.

\section{Data and Monte Carlo}

\par 
The data used for this analysis were collected by the L3 detector \cite{L3}
 at centre-of-mass energies \sqs {} = 189 $-$ 209 \GeV , with a
luminosity weighted  average 
value of \sqs {} = 198 \GeV, and a total integrated luminosity of 560 \pb .
Results on inclusive jet production at LEP for a smaller data sample
at lower \sqs {} were previously reported \cite{opal}.

\par
The  process ${\rm e}^{+} {\rm e}^{-} \rightarrow {\rm e}^{+} {\rm e}^{-}   hadrons$  
is modelled with the PYTHIA  \cite{PYTHIA} event generator 
with an event sample two times larger than the data.
In this generator, each photon can interact as a point-like particle (direct process), 
as a vector meson (VDM process) or
as a resolved photon (resolved process), leading to six classes of events. Since both incoming photons are assumed to be on the mass 
 shell, PYTHIA  is modified     to generate   the photon flux
 in the Equivalent Photon Approximation~\cite{epa}.
Predictions from the PHOJET \cite{Engel} Monte Carlo program are also compared with the data.
The following Monte Carlo generators are used to simulate the relevant 
background processes: KK2f\cite{KK2f} for 
\ee $\rightarrow \rm q \bar{q} \,(\gamma $); 
KORALZ \cite{KORALZ} for \ee $\rightarrow \tau^{+} \tau^{-}(\gamma )$;
KORALW \cite{KORALW} for \ee $\rightarrow \rm{W}^{+} \rm{W}^{-}$ 
and  DIAG36 \cite{DIAG36} for \ee \ra {} \ee $\tau^{+} \tau^{-}$.
Jet hadronisation is simulated with the JETSET  \cite{PYTHIA} parton shower algorithm.
Events are simulated in the L3 detector using the GEANT \cite{GEANT}
and GHEISHA \cite{GHEISHA} programs
and passed through the same reconstruction program as the data.
Time dependent detector inefficiencies, as monitored during each data taking
period, are included in the simulations.

\section{Event selection}

\par
\begin{figure}
    \includegraphics[width=0.95\linewidth]{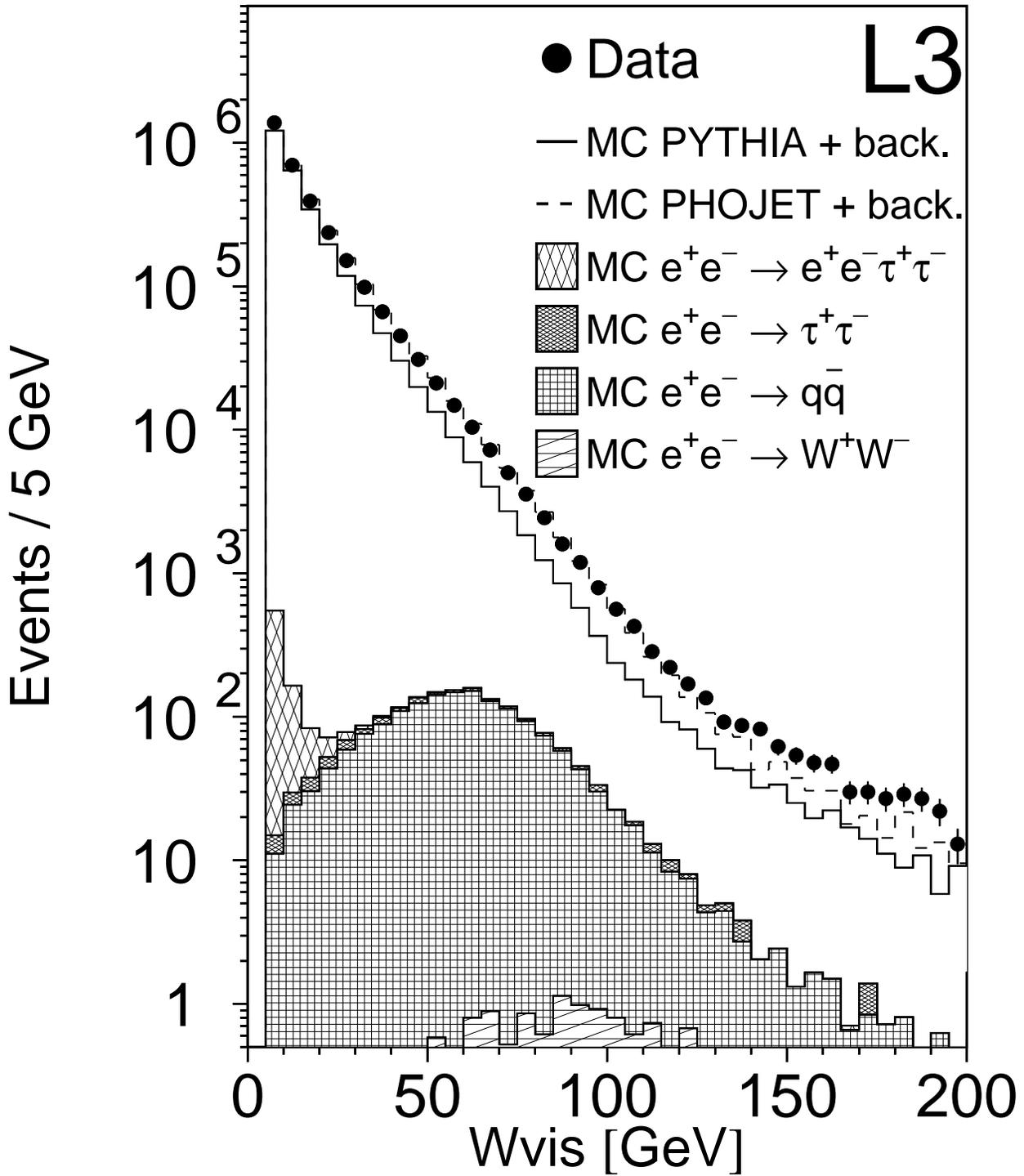}
  \caption{Distribution of the visible mass for selected events. The Monte Carlo distributions are normalised
  to the luminosity of the data.
  Various contributions to the background (back.) are shown as cumulative histograms.}
  \label{fig:wvis}
\end{figure}

Two-photon interaction events  are collected predominantly by the track triggers \cite{tracktrig}
with a low transverse momentum threshold of about 150 \MeV .
The selection of  ${\rm e}^{+} {\rm e}^{-} \rightarrow 
{\rm e}^{+} {\rm e}^{-} hadrons$ events \cite{l3tot} consists of:

\begin{itemize}
\item 
A multiplicity cut. To select hadronic final states, at  
least six objects must be detected,
where an object can be a track satisfying minimal quality requirements  
 or a calorimetric cluster of energy greater than 100 \MeV .
 
\item 
Energy cuts. 
To suppress background from beam-gas and beam-wall interactions, 
the total energy in the electromagnetic calorimeter is required to be 
greater than 500 \MeV .
In order to exclude \ee {} annihilation events, 
the total  energy deposited in the calorimeters must be 
less than 0.4 \sqs .

\item
An anti-tag condition. 
Events with a cluster in the luminosity monitor, which covers the angular region
$31<\theta< 62$ mrad,
with an electromagnetic shower shape and
energy greater than 30 \GeV {} are excluded. 

\item
A mass cut. The mass of all the visible particles of the event, including clusters in the luminosity monitor, must be greater than 5 \GeV .
In this calculation, the pion mass is attributed to tracks and electromagnetic clusters are 
treated as massless.
The visible mass distribution for data and Monte Carlo is shown, after all cuts, in Figure
\ref{fig:wvis}. A wide range of masses is accessible.

\end{itemize}

About 3 million  hadronic events are selected by these criteria.
The background level of this sample is less than  1\% and is 
mainly due to the $\ee \rightarrow \rm q \bar{q} \,(\gamma)$,
 \ee \ra {} $\tau^{+} \tau^{-}$ and
 \ee \ra {} \ee $\tau^{+} \tau^{-}$  processes.

\section{Jet definition and composition}

\par
Jets are formed from good quality tracks and electromagnetic 
clusters.
The tracks  have a
transverse momentum greater than 400 \MeV , an  
absolute pseudorapidity less than 1 and a distance of closest 
approach  to the primary vertex in the transverse plane less than 4 mm. 
The number of hits must be greater than 80\% of the maximum number expected 
from the track angle.   
For a transverse momentum less than 20 \GeV , the momentum and direction of the tracks are 
measured with the central tracker. 
For the tracks with transverse momentum above 20 \GeV , the track momenta are replaced with 
that derived from the energy of their associated cluster in the 
electromagnetic and hadronic calorimeters, assuming the pion mass.
Tracks associated with muon chamber hits are rejected.
An electromagnetic cluster must have an energy greater than 100 \MeV {} in 
at least 2 neighbouring
 BGO crystals and an absolute pseudorapidity less than 3.4. There should be no charged track 
within an angle of 200 mrad around the cluster direction and 
the associated energy  in the hadron calorimeter must be less
than 20\% of the electromagnetic energy.

\par
Jets are constructed using the $k_t$ jet algorithm KTCLUS \cite{Seymour}. 
This algorithm uses cylindrical geometry in which the distance between two objects 
 $i,\,j$ of transverse momenta 
$p_{ti}$ and $p_{tj}$ is
defined as $d_{ij}={\rm min}(p_{ti}^2,p_{tj}^2)[(\eta_i-\eta_j)^2+(\Phi_i-\Phi_j)^2]/D^2$ 
where $\eta_{i}$ and $\eta_{j}$ are the pseudorapidities of the objects,
$\Phi_{i}$ and $\Phi_{j}$ their azimuthal angles with respect to the beam axis
and $D$ is a parameter of the algorithm which determines the size of the jet. 
The standard value $D=1$ is used. 
A distance parameter $d_k$ equal to $p_{tk}^2$ is also associated to each object.
At the first iteration of the algorithm, the objects are the tracks and 
electromagnetic clusters defined above.
At each iteration of the algorithm, the  $d_{ij}$ and $d_k$ are ordered.
If the smallest distance is a $d_{ij}$, the corresponding 
objects $i$ and $j$ are replaced by a new object, a ``precluster'', formed by adding the 4-momenta 
of the objects $i$ and $j$.
If the smallest distance is a $d_k$ associated with a particle, this
is considered as a ``beam jet'' particle and is removed from the list of objects.
If the smallest distance is a $d_k$ associated with a precluster, this
defines a ``hard jet''  and is removed from the list of objects.
The procedure is iterated until all objects
define beam or hard jets. Only hard jets with $\pt>3$ \GeV {} and $|\eta|<1$ are 
further considered for this analysis.

\par
\begin{figure}
    \includegraphics[width=0.95\linewidth]{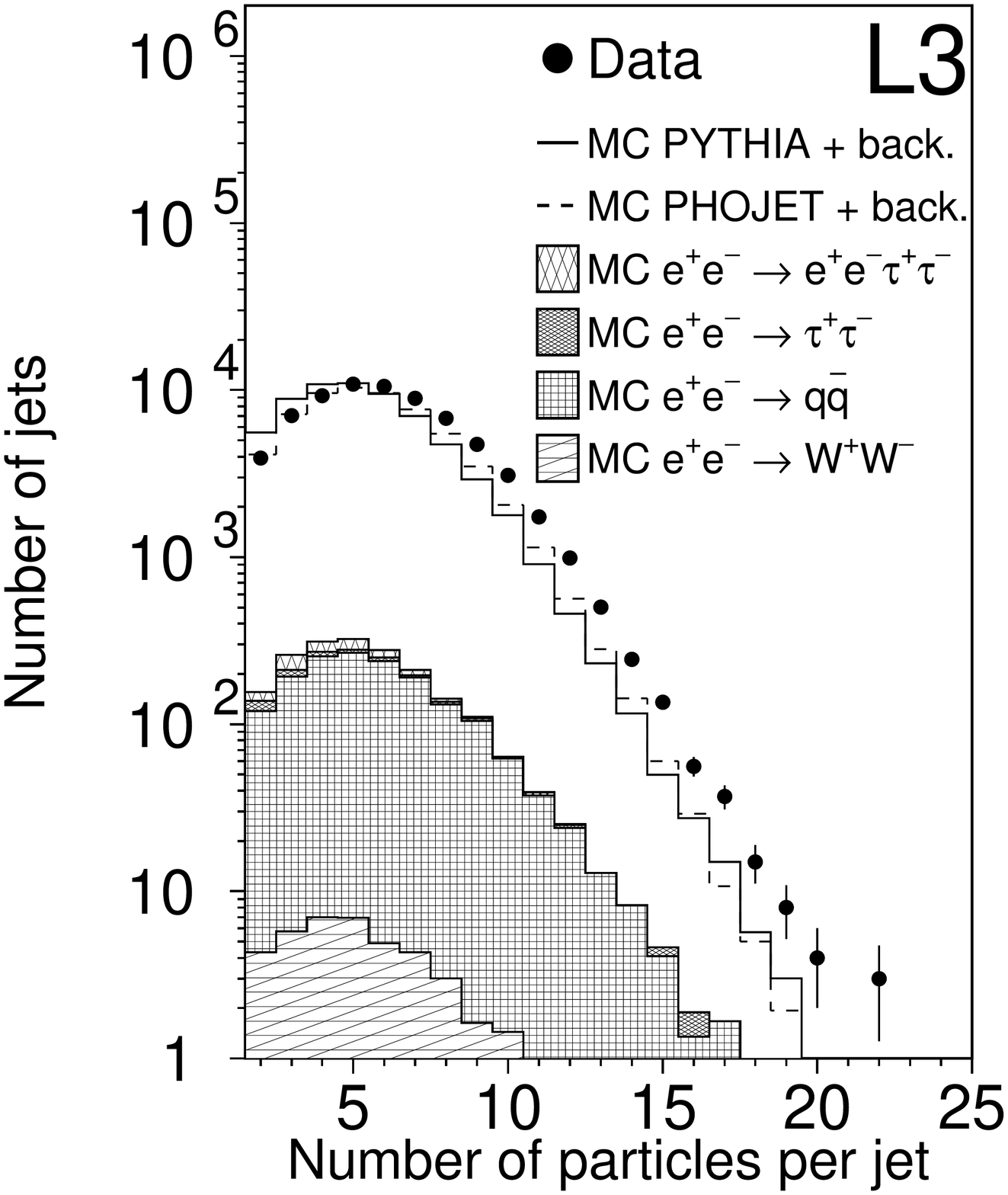}
  \caption{Distribution of the number of particles per jet for jets
  with $\pt >3$ \GeV {} and $|\eta|<1$. The Monte Carlo distributions  are normalised
  to the luminosity of the data.
  Various contributions to the background (back.) are shown as cumulative histograms.}
  \label{fig:npart}
\end{figure}

In Table \ref{tab:var}, the data are compared to the Monte Carlo at reconstructed and generated levels
for: the number of jets, the mean number of jets per event with at least one jet, the mean number of
particles per jet and outside the jets. For different \pt {} intervals, comparisons
are made of the mean number of tracks and electromagnetic clusters per jet and of transverse momentum 
of the leading particle 
divided by that of the jet. 
The standard deviations of these distributions are also quoted.
For Monte Carlo  at generator level, all particles with mean life time
less than $3\times10^{-10}$ s are allowed to decay and jets are formed from the  
photons, charged pions, 
charged and neutral kaons, protons and neutrons.
Both Monte Carlo programs underestimate the number of particles inside and outside the jets. 
The predicted number of electromagnetic clusters is too low for all \pt .
The amount of energy carried by the most energetic particle of the jet is correctly reproduced,
except in the highest \pt {} interval.  
The number of particles per jet is shown in Figure \ref{fig:npart}.
\par
\begin{figure}
    \includegraphics[width=0.45\linewidth]{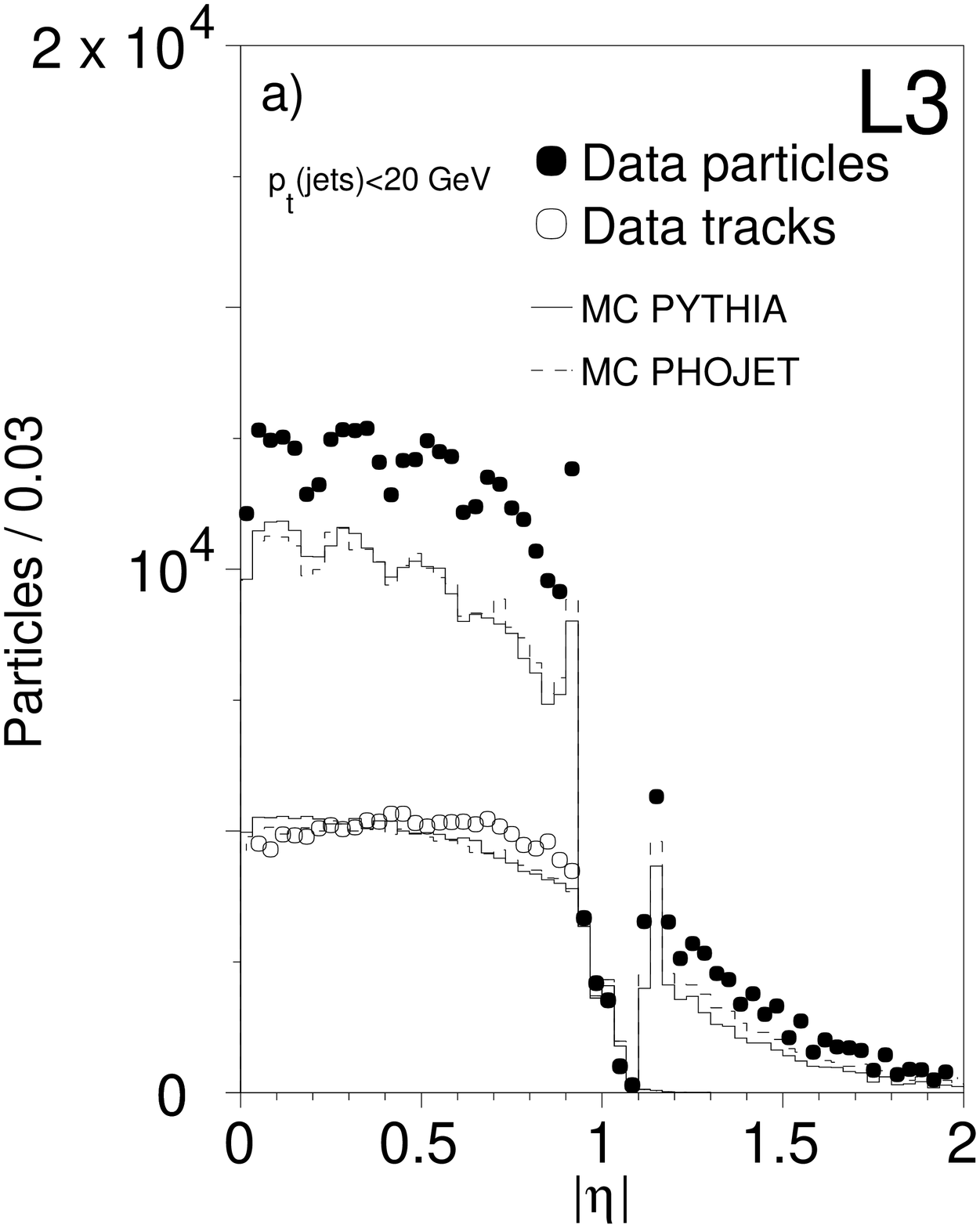}
    \includegraphics[width=0.45\linewidth]{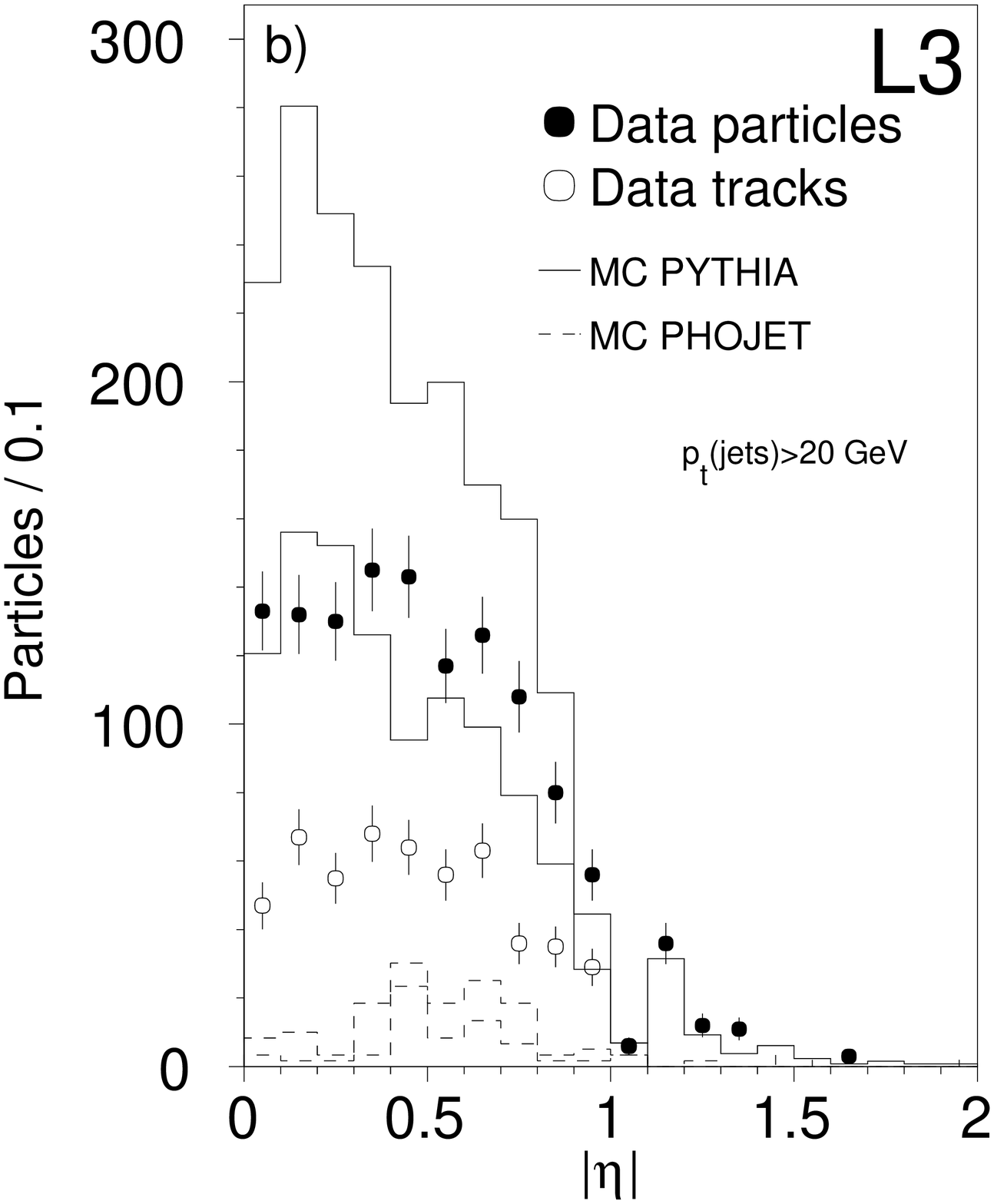} 
    \includegraphics[width=0.45\linewidth]{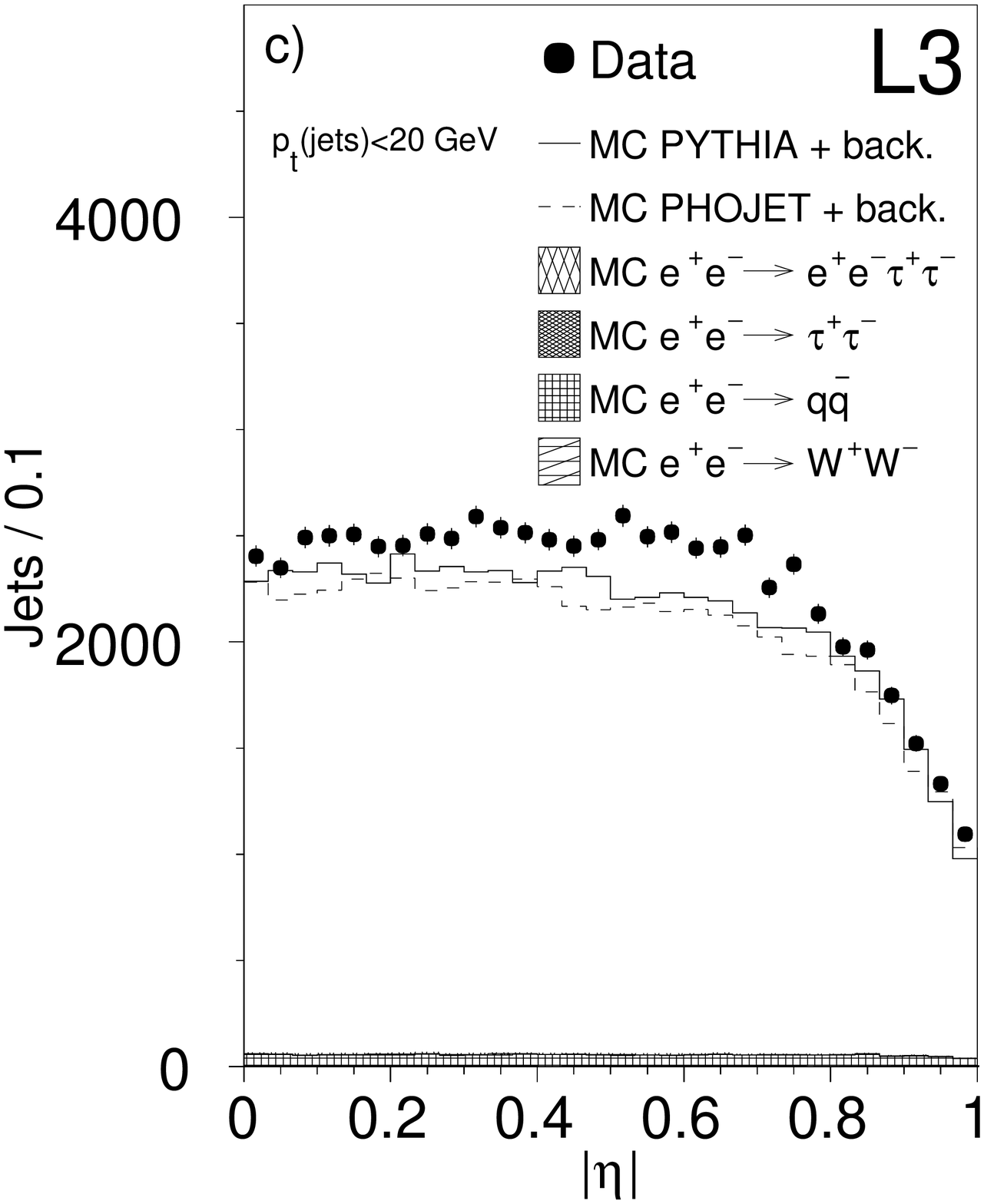}
    \includegraphics[width=0.45\linewidth]{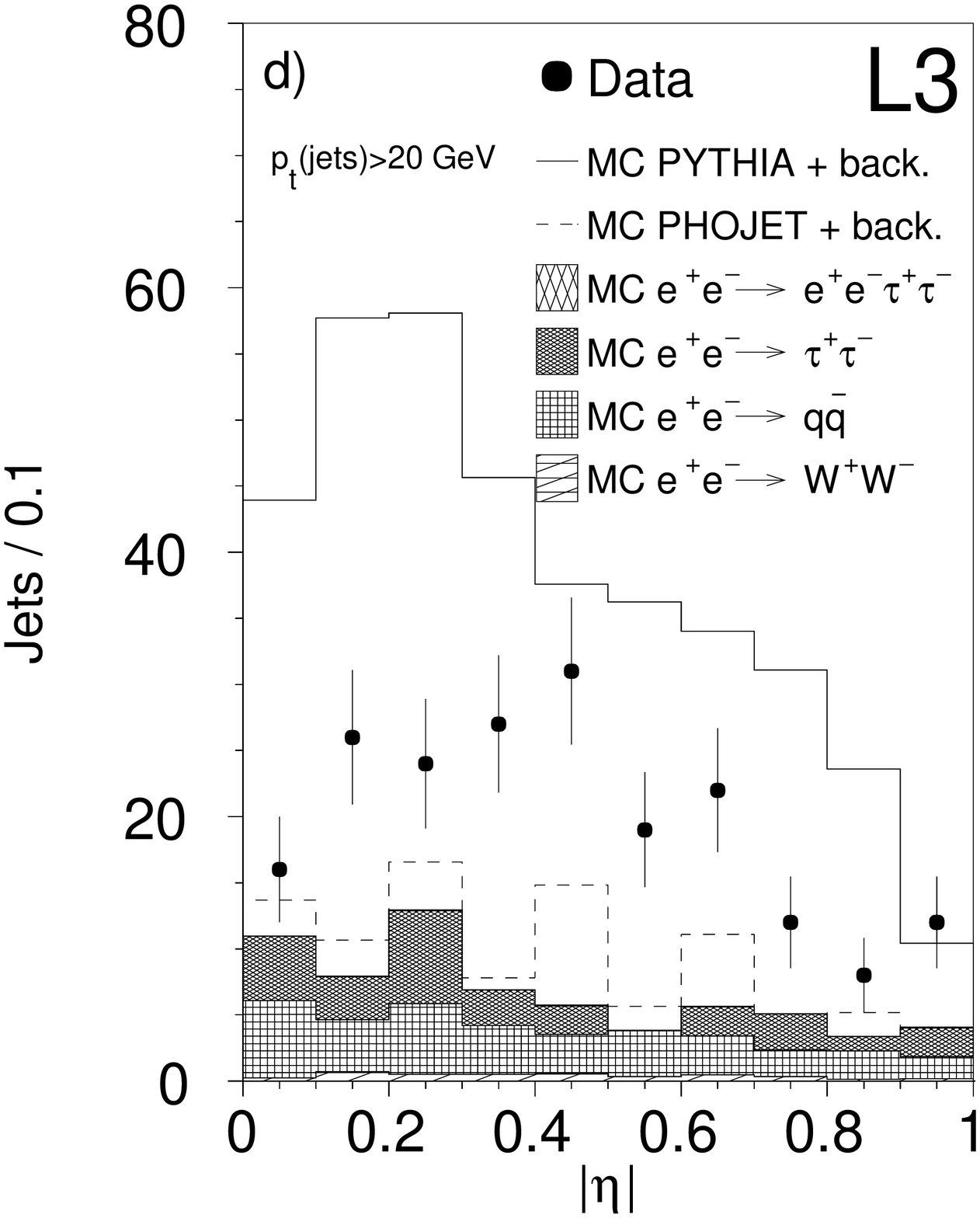} 
     
  \caption{Distributions of the pseudo rapidity $|\eta|$ for a) and b)
  particles and tracks used to form jets with $ \pt < 20 \GeV$ and $
  \pt \ge 20 \GeV$, respectively. ``Particles'' include both
  calorimetric clusters and tracks. c) and d) distributions of
  $|\eta|$ for reconstructed jets with $ \pt < 20 \GeV$ and $ \pt \ge
  20 \GeV$, respectively.  The Monte Carlo distributions are
  normalised to the luminosity of the data.  In a) and b) the higher Monte Carlo lines 
  refer to particles and the lower ones to tracks. Various contributions to
  the background are shown as cumulative histograms in c) and
  d).}
  \label{fig:eta}
\end{figure}

Figure \ref{fig:eta} shows the distributions of $|\eta|$ for
particles, {\it i.e.} clusters and tracks, tracks and jets in
two intervals of the jet transverse momentum, $ \pt < 20 \GeV $ and  $ \pt \ge 20 \GeV $. The detector acceptance for tracks,
 calorimetric clusters and jets is well reproduced by Monte Carlo models.

\section{Differential cross section}

\par
The differential cross section for inclusive jet production 
as a function of 
\pt {} is measured for 
$\Wgg \ge 5 \GeV $,
with a mean value of $\langle \Wgg \rangle \simeq 30 \GeV$,
and a photon virtuality $Q^2 < 8 ~ \GeV ^2$, with 
$\langle Q^2 \rangle \simeq 0.2 ~ \GeV ^2$.
This phase space is defined by Monte Carlo generator-level cuts.
Results are presented in 9 \pt {}
bins  between 3 and 50 \GeV .

\par
\begin{figure}
    \includegraphics[width=0.95\linewidth]{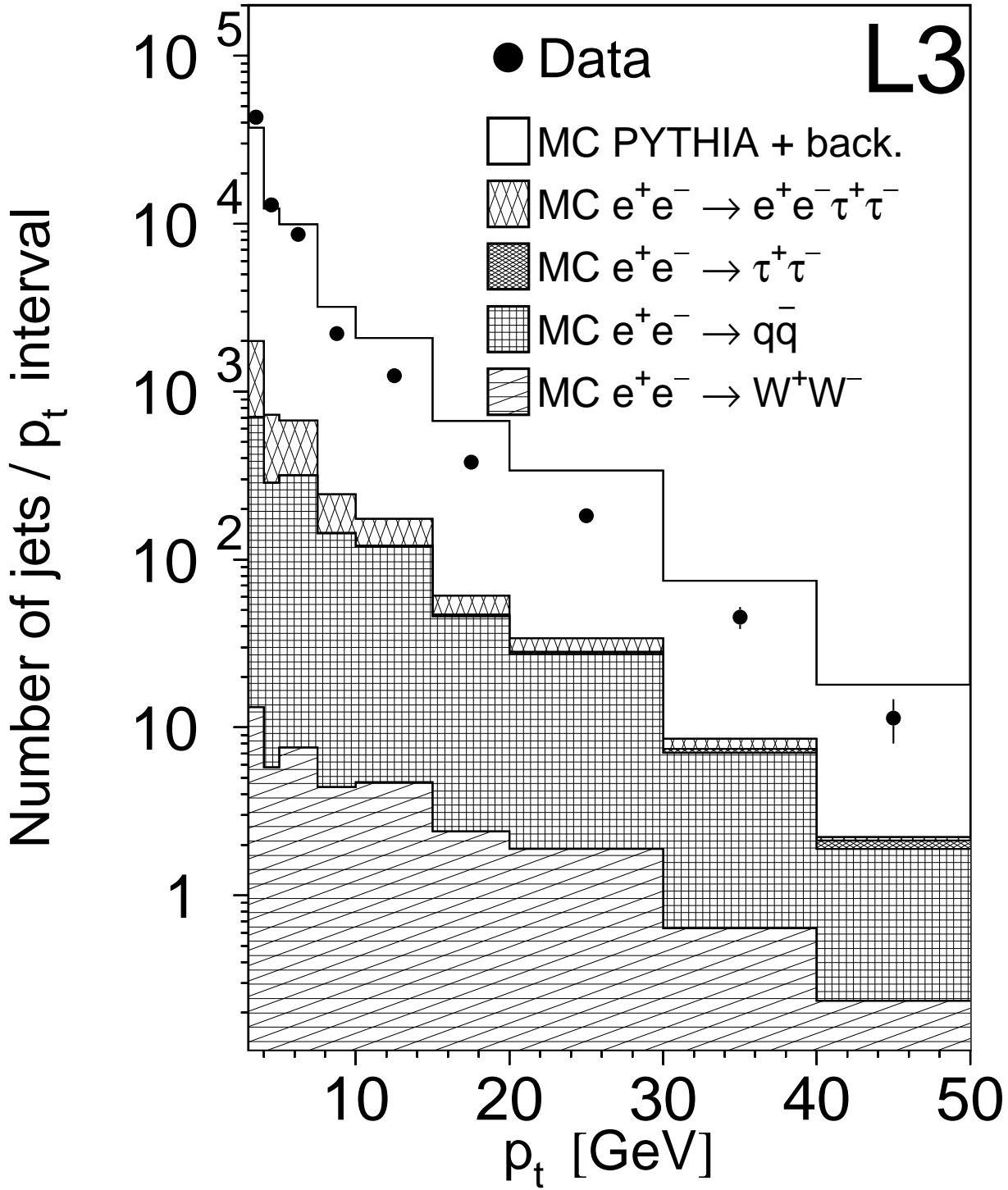}
  \caption{Distribution of the number of jets with $|\eta|<1$ as a function of \pt .
  The Monte Carlo distributions  are normalised
  to the luminosity of the data.
  Various contributions to the background (back.) are shown as cumulative histograms.}
  \label{fig:raw}
\end{figure}

The \pt {} distribution of the jets is presented
in Figure \ref{fig:raw}. The total  background
is listed in Table \ref{tab:jetpt}.
Events from the \ee \ra {} \ee $\tau^{+} \tau^{-}$  process dominate the background at low \pt {}
while hadronic and tau-pair annihilation events dominate it at high \pt .
To measure the cross section, the background is subtracted bin-by-bin.
The migration due to the \pt {} resolution
is corrected by a one-step Bayesian unfolding \cite{bayes}.
The data are  corrected for the selection efficiency which includes
acceptance, and is
calculated bin-by-bin as the ratio of the number of fully simulated
jets selected in PYTHIA over the number of generated jets, as formed
by the KTCLUS algorithm applied to particles at generator level. 
The efficiency decreases with \pt {} 
from  61\% to 15\%.

\par
The level 1 trigger efficiency is obtained
by comparing  the number of events accepted by 
the independent track and   
calorimetric energy triggers \cite{etrig}. It varies from 97\% to 100\%. 
The efficiency of higher level triggers is about 98\% and is measured using
prescaled  events. 
The differential cross section and the overall efficiency, which take into account selection and
trigger efficiencies, are given as a function of \pt {} in Table \ref{tab:jetpt}.

\par
Sources of systematic uncertainties on the cross section measurements
are the uncertainties on the estimation of the selection and trigger
efficiencies,
the limited Monte Carlo statistics, the background subtraction procedure, 
the selection procedure and the Monte Carlo modelling. Their contributions are shown in Table \ref{tab:err}.
The uncertainty due to the selection procedure is evaluated by 
repeating the analysis with different selection criteria:
the multiplicity cut is moved to 5 and to 7 objects, 
the requirement on the number of hits of the tracks is moved to 70\% of those expected, 
the isolation angle of clusters is moved to 100 mrad, 
and jets with a particle accounting for more than 90\% of the jet transverse momentum are rejected.
The sum in quadrature of the differences between these and the reference results
is assigned as systematic uncertainty in Table \ref{tab:err}.
Varying other criteria, such as the energy cut, the minimum cluster energy or the threshold 
where the track energy is defined by calorimeters,
gives negligible contributions. 
To evaluate the uncertainty on the Monte Carlo modelling,
the selection efficiency is determined using only one of the PYTHIA subprocesses: VDM-VDM,
direct-direct or resolved-resolved. 
The systematic uncertainty is assigned as the maximum difference between these values and the
reference Monte Carlo.
\par
The differential cross sections as a function of $|\eta |$ are uniform within the experimental uncertainties
for both $\pt < 20 \GeV$ and  $\pt > 20 \GeV$, albeit in the latter case these uncertainties are large.

\par
\begin{figure}
   \includegraphics[width=0.5\linewidth]{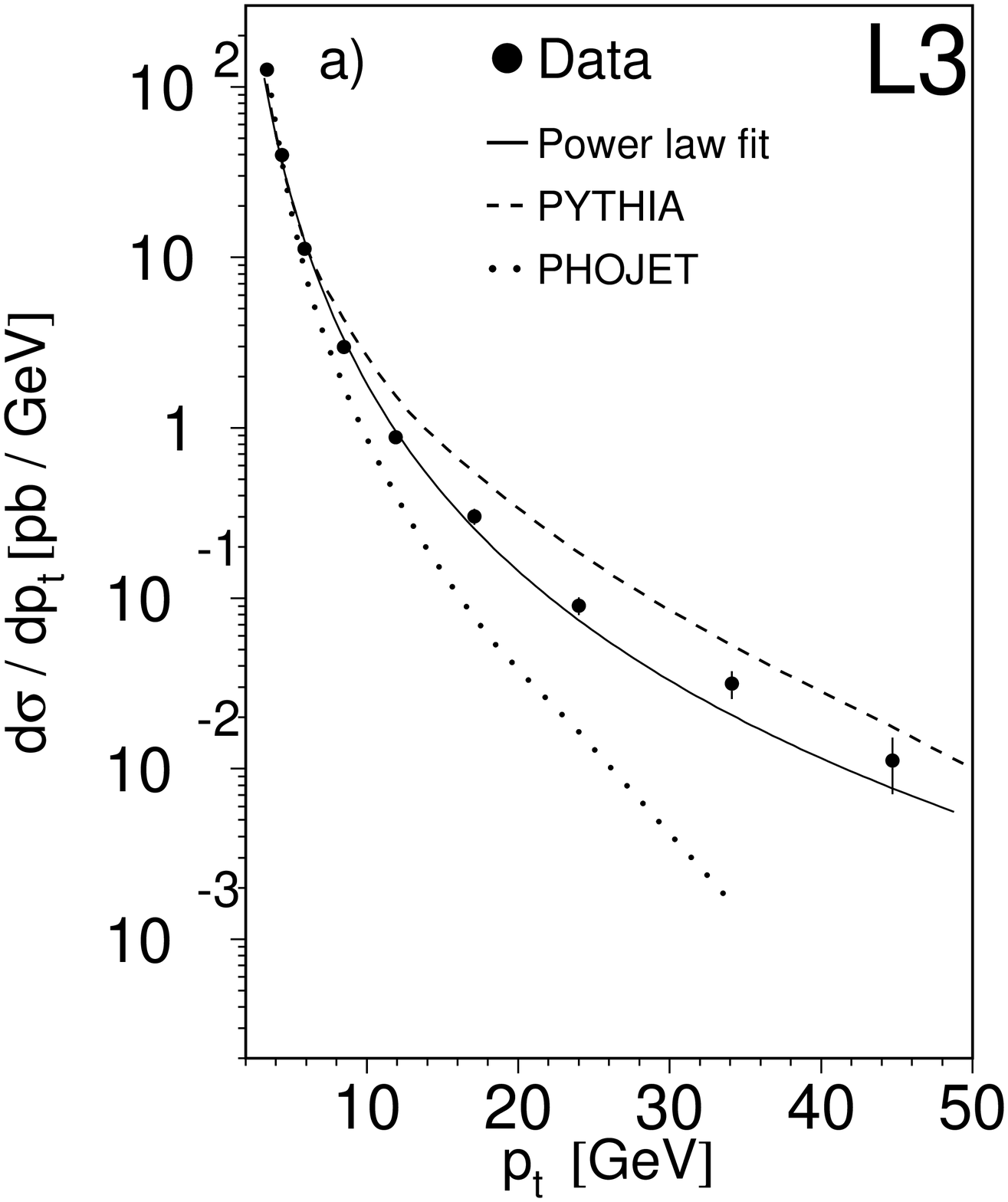}
    \includegraphics[width=0.5\linewidth]{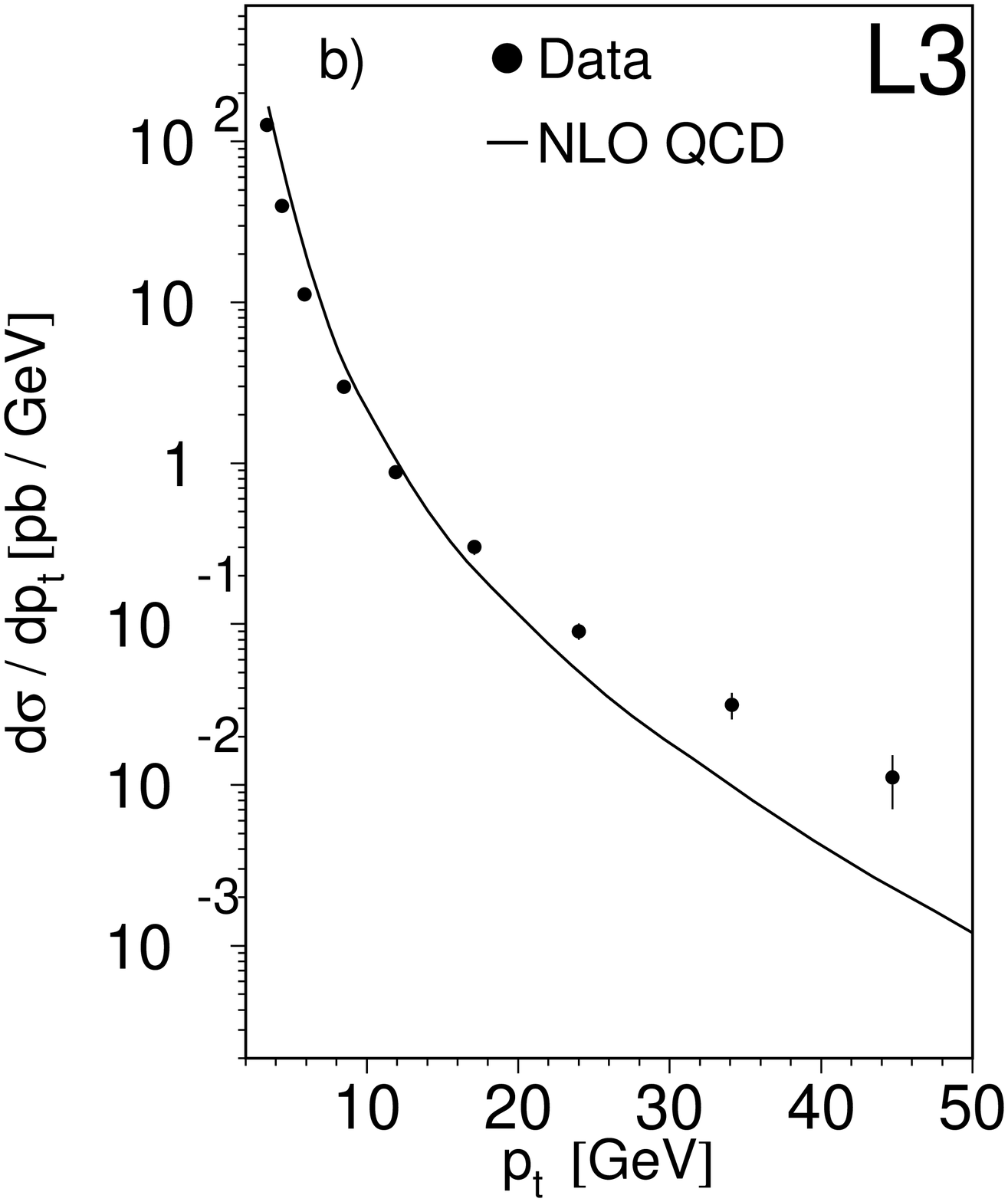}
  \caption{ Inclusive jet differential cross section \dpt {} 
 a)  compared to PYTHIA and PHOJET Monte Carlo predictions 
   and the result of a power law fit
  (solid line);
  b) compared to NLO QCD calculations [2] (solid line).
  The theoretical scale uncertainty is less than 20\%.}
  \label{fig:qcd}
\end{figure}

The  differential cross section \dpt {} is described 
by a power law function $A \pt^{-B}\!$,
as expected from the onset of hard QCD processes, with $B =  3.65\pm0.07$. 
The result of the fit is shown in Figure \ref{fig:qcd}a together with a comparison to Monte Carlo
predictions.

\par
\begin{figure}
    \includegraphics[width=0.95\linewidth]{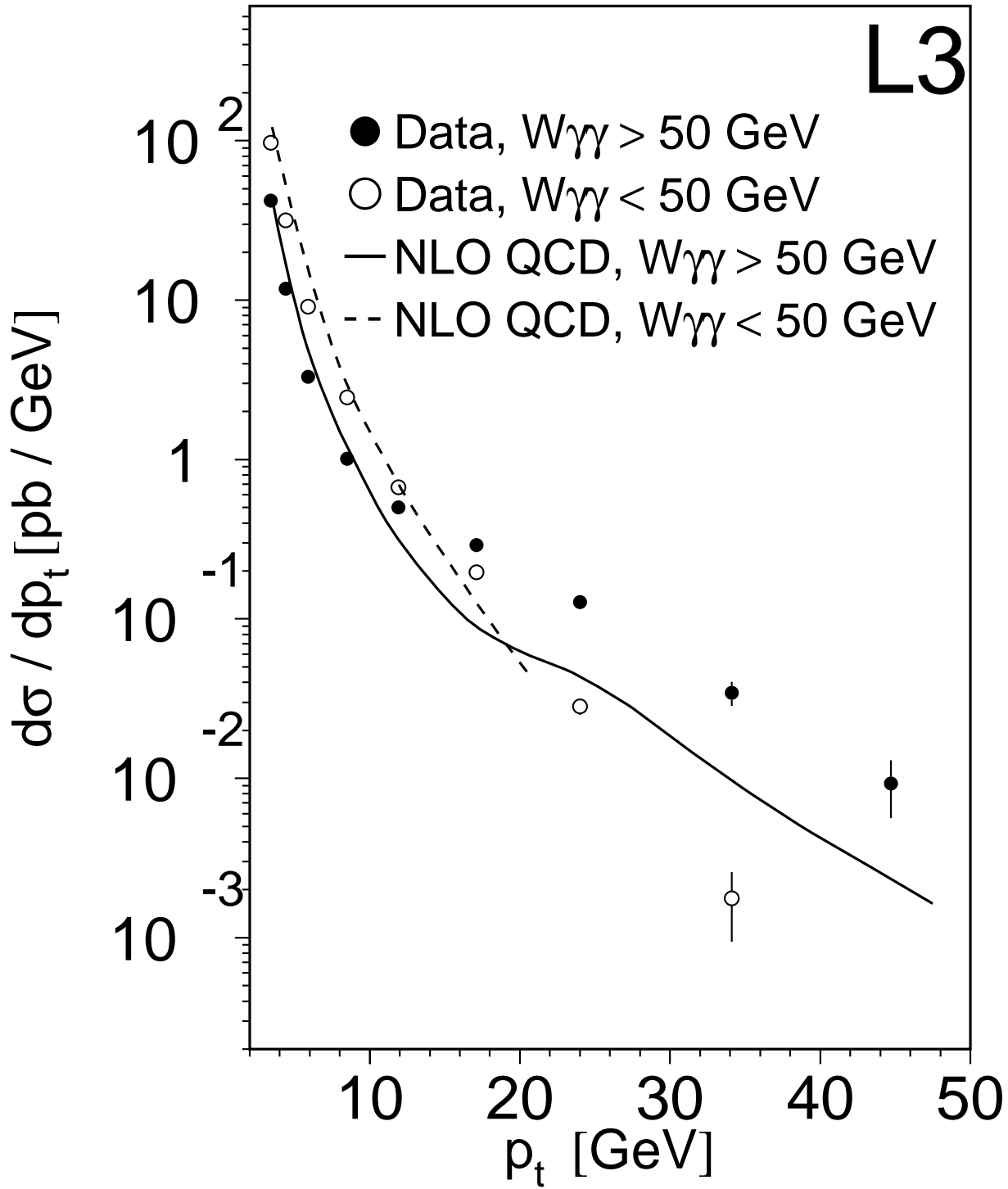}
  \caption{ Inclusive jet differential cross section \dpt {} 
  for events with two-photon centre-of-mass energy, \Wgg , 
  below and above 50 \GeV . NLO QCD calculations [2] 
  are superimposed to the data. The discontinuity around $25\GeV$ is
  due to the direct contribution.}
  \label{fig:masscut}
\end{figure}

In Figure \ref{fig:qcd}b the data are also compared
to  analytical NLO QCD predictions \cite{frixione2}. 
For this calculation,
the flux of quasi-real  photons is obtained using
the improved Weizs\"acker-Williams formula \cite{WW}. 
The interacting particles can be point-like photons or 
partons  from the $\gamma \ra \rm{q\bar{q}}$ process, which
evolve into quarks and gluons.
The GRV-HO parton density functions of
Reference  \citen{GRV} are used and
all elementary $2 \ra 2$ and $2\ra 3$ processes  are considered. 
The parameter $\Lambda^{(5)}$ is set to 130 \MeV .
The renormalization and factorisation scales
are taken to be equal: $\mu=M=E_t/2$ \cite{frixione}.
To assign uncertainties, the scale is varied by a factor 1/2 or 2, 
which gives a change in the prediction less than 20\%. 
The results of this calculation agree~\cite{frixione2} with those described in 
Reference~\citen{kramer}.
An additional uncertainty
in comparison with NLO QCD, which is not considered here, might arise from the modeling 
of the hadronisation process. In a similar study~\cite{opalTwoJets} it was evaluated to be below 10\% for $\pt>10\GeV$ and
decreasing with increasing $\pt$. 
The agreement with the data is poor in the high-\pt {} range, as previously
observed in the case of inclusive \pz {}  \cite{pz} and \ppm {} \cite{ch}
production in similar two-photon reactions. In Figure \ref{fig:masscut}, the data are
divided in two \Wgg {} ranges, $\Wgg > 50 \GeV$ and $\Wgg \ge 50 \GeV$
and compared to the analytical NLO QCD predictions\cite{frixione2}. 
For $\Wgg \ge 50 \GeV$ there is 
a clear  discontinuity in the slope
near  $\pt = 25 \GeV$, 
  due to the direct  contribution. At  high \pt , the disagreement between data
  and theoretical calculations is still present.

\newpage
\section*{Author List}
\input namelist274.tex

\begin{sidewaystable}
  \begin{center}

\begin{tabular}{|l@{~~}r||r@{~$\pm$~}l@{~}l|r@{~}l|r@{~}l|r@{~}l|r@{~}l|}
\hline
\multicolumn{2}{|l||}{Variable} & \multicolumn{3}{|c|}{Data} & \multicolumn{4}{|c|}{PYTHIA} & \multicolumn{4}{|c|}{PHOJET}  \\
\cline{6-13}
\multicolumn{2}{|l||}{}&\multicolumn{3}{|c|}{}& \multicolumn{2}{|c|}{Reconstructed} & \multicolumn{2}{|c|}{Generated} & \multicolumn{2}{|c|}{Reconstructed} & \multicolumn{2}{|c|}{Generated} \\
\hline
\hline
\multicolumn{2}{|l||}{Total number of jets}  & \multicolumn{3}{|c|}{68792} & \multicolumn{2}{|c|}{107140} &  \multicolumn{2}{|c|}{188302} &  \multicolumn{2}{|c|}{65781} &   \multicolumn{2}{|c|}{105633} \\
\multicolumn{2}{|l||}{Number of jets / event}  &   1.2\phantom{0}&0.1&(0.5) &   \phantom{00}1.4\phantom{0}&(0.7) &   \phantom{0}1.3\phantom{0}&(0.7) &   \phantom{00}1.2\phantom{0}&(0.4) &   \phantom{0}1.1\phantom{0}&(0.5) \\
\hline
\multicolumn{2}{|l||}{N(particles) / jet}        &  6.1\phantom{0}&0.1&(2.5) &  \phantom{00}5.4\phantom{0}&(2.3) &   \phantom{0}5.3\phantom{0}&(2.4) &   \phantom{00}5.7\phantom{0}&(2.4) &   \phantom{0}6.1\phantom{0}&(2.4) \\
\multicolumn{2}{|l||}{N(particles) outside jets}   &  14.4\phantom{0}&0.1&(8.4) &  \phantom{0}10.0\phantom{0}&(7.0) &  13.6\phantom{0}&(9.3) &  \phantom{0}12.4\phantom{0}&(7.3) &  18.4\phantom{0}&(8.8)  \\
\hline
N(tracks) / jet &  3 $\!<\!\pt\!<\!$ 5\phantom{0} \GeV       &  2.2\phantom{0}&0.1&(1.3) &   \phantom{00}2.3\phantom{0}&(1.3) &   \multicolumn{2}{|c|}{ } &   \phantom{00}2.4\phantom{0}&(1.3) &   \multicolumn{2}{|c|}{ } \\
                   &   5 $\!<\!\pt\!<\!$ 10 \GeV           &  2.4\phantom{0}&0.1&(1.3) &   \phantom{00}2.6\phantom{0}&(1.3) &   \multicolumn{2}{|c|}{ } &   \phantom{00}2.8\phantom{0}&(1.4) &   \multicolumn{2}{|c|}{ } \\
                   &   10 $\!<\!\pt\!<\!$ 25 \GeV          &  2.5\phantom{0}&0.1&(1.6) &   \phantom{00}2.9\phantom{0}&(1.3) &   \multicolumn{2}{|c|}{ } &   \phantom{00}3.0\phantom{0}&(1.6) &   \multicolumn{2}{|c|}{ } \\
                   &   25 $\!<\!\pt\!<\!$ 45 \GeV          &  2.7\phantom{0}&0.2&(1.7) &   \phantom{00}3.3\phantom{0}&(1.6) &   \multicolumn{2}{|l|}{  
\begin{picture}(10,10)
\put (-6.4,-4.6){\line(5,4){73}}
\put (-6.4,54){\line(5,-4){73}}
\end{picture}
}		   &  \multicolumn{2}{|c|}{$-$}           &  \multicolumn{2}{|l|}{
\begin{picture}(10,10)
\put (-6.4,-4.6){\line(5,4){73}}
\put (-6.4,54){\line(5,-4){73}}
\end{picture}
} \\
\hline
N(clusters) / jet  &  3 $\!<\!\pt\!<\!$ 5\phantom{0} \GeV    &  3.7\phantom{0}&0.1&(2.4) &   \phantom{00}2.0\phantom{0}&(2.0) &   \multicolumn{2}{|c|}{ } &  \phantom{00}3.1\phantom{0}&(2.2) &   \multicolumn{2}{|c|}{ } \\
                   &  5 $\!<\!\pt\!<\!$ 10 \GeV            &  3.9\phantom{0}&0.1&(2.6) &   \phantom{00}1.8\phantom{0}&(1.9) &   \multicolumn{2}{|c|}{ } &   \phantom{00}3.3\phantom{0}&(2.4) &   \multicolumn{2}{|c|}{ } \\
                   &  10 $\!<\!\pt\!<\!$ 25 \GeV           &  3.9\phantom{0}&0.1&(3.0) &   \phantom{00}1.6\phantom{0}&(1.8) &   \multicolumn{2}{|c|}{ } &   \phantom{00}3.3\phantom{0}&(2.5) &   \multicolumn{2}{|c|}{ } \\
                   &  25 $\!<\!\pt\!<\!$ 45 \GeV           &  3.8\phantom{0}&0.3&(3.0) &   \phantom{00}1.4\phantom{0}&(1.7) &   \multicolumn{2}{|l|}{  
\begin{picture}(10,10)
\put (-6.4,-4.6){\line(5,4){73}}
\put (-6.4,54){\line(5,-4){73}}
\end{picture}
 } &  \multicolumn{2}{|c|}{$-$}           &  \multicolumn{2}{|l|}{  
\begin{picture}(10,10)
\put (-6.4,-4.6){\line(5,4){73}}
\put (-6.4,54){\line(5,-4){73}}
\end{picture}
 } \\
\hline
\pt (leading) / \pt  &  3 $\!<\!\pt\!<\!$ 5\phantom{0} \GeV   &  0.50&0.01&(0.18) &  \phantom{0}0.53&(0.18) &  0.50&(0.18) &  \phantom{0}0.51&(0.18) &   0.46&(0.17) \\
                    &  5 $\!<\!\pt\!<\!$ 10 \GeV           &  0.54&0.01&(0.20) &  \phantom{0}0.55&(0.19) &  0.50&(0.20) &  \phantom{0}0.52&(0.19) &   0.43&(0.17) \\
                    &  10 $\!<\!\pt\!<\!$ 25 \GeV          &  0.63&0.01&(0.23) &  \phantom{0}0.60&(0.20) &  0.48&(0.22) &  \phantom{0}0.60&(0.24) &   0.39&(0.19)  \\
                    &  25 $\!<\!\pt\!<\!$ 45 \GeV          &  0.69&0.03&(0.23) &  \phantom{0}0.56&(0.14) &  0.47&(0.25) &   \multicolumn{2}{|c|}{$-$} &  \multicolumn{2}{|c|}{$-$} \\
\hline

\end{tabular}

\caption{Mean value and standard deviation (in brackets) of multiplicities and \pt {} fractions 
for the jets in data and Monte Carlo events, at generator level as well as after reconstruction. 
The uncertainties on the mean values are quoted for the data. For Monte Carlo, they are
always lower than the precision of the last digit.
}
\label{tab:var}
  \end{center}
\end{sidewaystable}

\begin{table}
  \begin{center}

    \begin{tabular}{|r@{$-$}l|c||r@{~$\pm$~}l|r@{~$\pm$~}l|r@{~$\pm$~}l|r@{~$\pm$~}r@{~$\pm$~}lc|}
    \hline
    \multicolumn{2}{|c|}{\pt}  & $\langle\pt\rangle$  & \multicolumn{2}{|c|}{Background} & \multicolumn{2}{|c|}{Reconstruction} & \multicolumn{2}{|c|}{Trigger} & \multicolumn{4}{|c|}{\dpt} \\
     \multicolumn{2}{|c|}{[\GeV]}  & [\GeV] & \multicolumn{2}{|c|}{[\%]} & \multicolumn{2}{|c|}{efficiency [\%]} & \multicolumn{2}{|c|}{efficiency [\%]} & \multicolumn{4}{|c|}{ [pb/\GeV] }\\
    \hline
    3&4     & 3.4   & \phantom{00}4.6 & 0.1\phantom{0} & \phantom{00}60.8 & 0.2\phantom{00} & \phantom{0}95.8 & 0.3\phantom{0} & \phantom{0}(13 & 1 & 1) & $\times 10^{1}$     \\
    4&5     & 4.4   & \phantom{0}5.6 & 0.1 & 57.2 & 0.3 & 95.9 & 0.5 & (40 & 1 & 3) &   \\
    5&7.5   & 5.9   & \phantom{0}7.8 & 0.1 & 53.2 & 0.3 & 96.2 & 0.5 & (11 & 1 & 1) &     \\
    7.5&10  & 8.5   & \phantom{}11.1 & 0.1 & 48.9 & 0.5 & 96.6 & 1.0 & (30 & 1 & 2) &$\times 10^{-1}$     \\
    10&15   & 11.9  & \phantom{}14.0 & 0.2 & 44.9 & 0.6 & 96.8 & 1.4 & (88 & 3 & 7) &$\times 10^{-2}$     \\
    15&20   & 17.1  & \phantom{}16.0 & 0.4 & 39.2 & 0.9 & 96.9 & 2.0 & (30 & 2 & 3) &$\times 10^{-2}$       \\
    20&30   & 24.0  & \phantom{}18.6 & 0.8 & 31.6 & 0.8 & 97.3 & 2.1 & (90 & 7 & 8) &$\times 10^{-3}$     \\
    30&40   & 34.1  & \phantom{}18.9 & 1.5 & 20.5 & 1.3 & 97.3 & 2.5 & (31 & 5 & 2) &$\times 10^{-3}$     \\
    40&50   & 44.7  & \phantom{}19.6 & 1.6 & 15.2 & 1.9 & 98.5 & 2.8 & (11 & 3 & 2) &$\times 10^{-3}$   \\
    \hline 
   \end{tabular}

    \caption{Background level, reconstruction efficiency, trigger efficiency and differential cross section as a function of \pt {} 
    for $|\eta| <$ 1 and $\Wgg >$ 5 \GeV.
    The first uncertainty is statistical and the second systematic.
    The average value of \pt {} for each bin, $\langle\pt\rangle$, is also given.} 
    \label{tab:jetpt}

  \end{center}
\end{table}

\begin{table}
  \begin{center}

    \begin{tabular}{|r@{$-$}l||c|c|c|c|c|}
    \hline
    \multicolumn{2}{|c||}{\pt     } & Trigger       &  Monte Carlo & Background       & Selection      &  Monte Carlo   \\
    \multicolumn{2}{|c||}{ [\GeV] } & efficiency [\%]& statistics [\%] & subtraction [\%] & procedure [\%] & modelling [\%]   \\
    \hline
       3&4    & 0.3 & \phantom{0}0.3& \phantom{}$<$ 0.1 \phantom{$<$} & \phantom{$<$} 8.4 \phantom{$<$} & \phantom{0}0.3   \\
       4&5    & 0.5 & \phantom{0}0.5 & \phantom{$<$} 0.2 \phantom{$<$} & \phantom{$<$} 7.0 \phantom{$<$} & \phantom{0}1.3  \\
       5&7.5  & 0.5 & \phantom{0}0.5& \phantom{$<$} 0.3 \phantom{$<$} & \phantom{$<$} 6.6 \phantom{$<$} & \phantom{0}1.5   \\
       7.5&10 & 1.0 & \phantom{0}1.0 & \phantom{$<$} 0.6 \phantom{$<$} & \phantom{$<$} 4.8 \phantom{$<$} & \phantom{0}2.4  \\
       10&15  & 1.4 & \phantom{0}1.3& \phantom{$<$} 0.9 \phantom{$<$} & \phantom{$<$} 7.0 \phantom{$<$} & \phantom{0}2.6   \\
       15&20  & 2.1 & \phantom{0}2.4& \phantom{$<$} 1.7 \phantom{$<$} & \phantom{$<$} 8.0 \phantom{$<$} & \phantom{0}3.3   \\
       20&30  & 2.2 & \phantom{0}2.6& \phantom{$<$} 2.7 \phantom{$<$} & \phantom{$<$} 6.0 \phantom{$<$} & \phantom{0}4.8   \\
       30&40  & 2.6 & \phantom{0}6.4& \phantom{$<$} 5.2 \phantom{$<$} & \phantom{}$<$ 0.1 \phantom{$<$} & \phantom{0}6.2   \\
       40&50  & 2.8 & \phantom{}12.4& \phantom{$<$} 9.6 \phantom{$<$} & \phantom{}$<$ 0.1 \phantom{$<$} & \phantom{}12.4   \\
                                                                                                                
    \hline
    \end{tabular}
    \caption{Systematic uncertainties on the inclusive jet cross section as a function of \pt .}
    \label{tab:err}

  \end{center}
\end{table}

\newpage

\end{document}

%% file: namelist274.tex
\typeout{   }     
\typeout{Using author list for paper 274 - 277 }
\typeout{$Modified: Jul 15 2001 by smele $}
\typeout{!!!!  This should only be used with document option a4p!!!!}
\typeout{   }
%
%
%
%
%
%

\newcount\tutecount  \tutecount=0
\def\tutenum#1{\global\advance\tutecount by 1 \xdef#1{\the\tutecount}}
\def\tute#1{$^{#1}$}
\tutenum\aachen            
\tutenum\nikhef            
\tutenum\mich              
\tutenum\lapp              
\tutenum\basel             
\tutenum\lsu               
\tutenum\beijing           
\tutenum\bologna           
\tutenum\tata              
\tutenum\ne                
\tutenum\bucharest         
\tutenum\budapest          
\tutenum\mit               
\tutenum\panjab            
\tutenum\debrecen          
\tutenum\dublin            
\tutenum\florence          
\tutenum\cern              
\tutenum\wl                
\tutenum\geneva            
\tutenum\hefei             
\tutenum\lausanne          
\tutenum\lyon              
\tutenum\madrid            
\tutenum\florida           
\tutenum\milan             
\tutenum\moscow            
\tutenum\naples            
\tutenum\cyprus            
\tutenum\nymegen           
\tutenum\caltech           
\tutenum\perugia           
\tutenum\peters            
\tutenum\cmu               
\tutenum\potenza           
\tutenum\prince            
\tutenum\riverside         
\tutenum\rome              
\tutenum\salerno           
\tutenum\ucsd              
\tutenum\sofia             
\tutenum\korea             
\tutenum\purdue            
\tutenum\psinst            
\tutenum\zeuthen           
\tutenum\eth               
\tutenum\hamburg           
\tutenum\taiwan            
\tutenum\tsinghua          

{
\parskip=0pt
\noindent
{\bf The L3 Collaboration:}
\ifx\selectfont\undefined
 \baselineskip=10.8pt
 \baselineskip\baselinestretch\baselineskip
 \normalbaselineskip\baselineskip
 \ixpt
\else
 \fontsize{9}{10.8pt}\selectfont
\fi
\medskip
\tolerance=10000
\hbadness=5000
\raggedright
\hsize=162truemm\hoffset=0mm
\def\r{\rlap,}
\noindent

P.Achard\r\tute\geneva\ 
O.Adriani\r\tute{\florence}\ 
M.Aguilar-Benitez\r\tute\madrid\ 
J.Alcaraz\r\tute{\madrid}\ 
G.Alemanni\r\tute\lausanne\
J.Allaby\r\tute\cern\
A.Aloisio\r\tute\naples\ 
M.G.Alviggi\r\tute\naples\
H.Anderhub\r\tute\eth\ 
V.P.Andreev\r\tute{\lsu,\peters}\
% ...
\newpage
\begin{list}{A}{\itemsep=0pt plus 0pt minus 0pt\parsep=0pt plus 0pt minus 0pt
                \topsep=0pt plus 0pt minus 0pt}
\item[\aachen]
 III. Physikalisches Institut, RWTH, D-52056 Aachen, Germany$^{\S}$
\item[\nikhef] National Institute for High Energy Physics, NIKHEF, 
     and University of Amsterdam, NL-1009 DB Amsterdam, The Netherlands
\item[\mich] University of Michigan, Ann Arbor, MI 48109, USA
\item[\lapp] Laboratoire d'Annecy-le-Vieux de Physique des Particules, 
     LAPP,IN2P3-CNRS, BP 110, F-74941 Annecy-le-Vieux CEDEX, France
\item[\basel] Institute of Physics, University of Basel, CH-4056 Basel,
     Switzerland
\item[\lsu] Louisiana State University, Baton Rouge, LA 70803, USA
\item[\beijing] Institute of High Energy Physics, IHEP, 
  100039 Beijing, China$^{\triangle}$ 
\item[\bologna] University of Bologna and INFN-Sezione di Bologna, 
     I-40126 Bologna, Italy
\item[\tata] Tata Institute of Fundamental Research, Mumbai (Bombay) 400 005, India
\item[\ne] Northeastern University, Boston, MA 02115, USA
\item[\bucharest] Institute of Atomic Physics and University of Bucharest,
     R-76900 Bucharest, Romania
\item[\budapest] Central Research Institute for Physics of the 
     Hungarian Academy of Sciences, H-1525 Budapest 114, Hungary$^{\ddag}$
\item[\mit] Massachusetts Institute of Technology, Cambridge, MA 02139, USA
\item[\panjab] Panjab University, Chandigarh 160 014, India.
\item[\debrecen] KLTE-ATOMKI, H-4010 Debrecen, Hungary$^\P$
\item[\dublin] Department of Experimental Physics,
  University College Dublin, Belfield, Dublin 4, Ireland
\item[\florence] INFN Sezione di Firenze and University of Florence, 
     I-50125 Florence, Italy
\item[\cern] European Laboratory for Particle Physics, CERN, 
     CH-1211 Geneva 23, Switzerland
\item[\wl] World Laboratory, FBLJA  Project, CH-1211 Geneva 23, Switzerland
\item[\geneva] University of Geneva, CH-1211 Geneva 4, Switzerland
\item[\hefei] Chinese University of Science and Technology, USTC,
      Hefei, Anhui 230 029, China$^{\triangle}$
\item[\lausanne] University of Lausanne, CH-1015 Lausanne, Switzerland
\item[\lyon] Institut de Physique Nucl\'eaire de Lyon, 
     IN2P3-CNRS,Universit\'e Claude Bernard, 
     F-69622 Villeurbanne, France
\item[\madrid] Centro de Investigaciones Energ{\'e}ticas, 
     Medioambientales y Tecnol\'ogicas, CIEMAT, E-28040 Madrid,
     Spain${\flat}$ 
\item[\florida] Florida Institute of Technology, Melbourne, FL 32901, USA
\item[\milan] INFN-Sezione di Milano, I-20133 Milan, Italy
\item[\moscow] Institute of Theoretical and Experimental Physics, ITEP, 
     Moscow, Russia
\item[\naples] INFN-Sezione di Napoli and University of Naples, 
     I-80125 Naples, Italy
\item[\cyprus] Department of Physics, University of Cyprus,
     Nicosia, Cyprus
\item[\nymegen] University of Nijmegen and NIKHEF, 
     NL-6525 ED Nijmegen, The Netherlands
\item[\caltech] California Institute of Technology, Pasadena, CA 91125, USA
\item[\perugia] INFN-Sezione di Perugia and Universit\`a Degli 
     Studi di Perugia, I-06100 Perugia, Italy   
\item[\peters] Nuclear Physics Institute, St. Petersburg, Russia
\item[\cmu] Carnegie Mellon University, Pittsburgh, PA 15213, USA
\item[\potenza] INFN-Sezione di Napoli and University of Potenza, 
     I-85100 Potenza, Italy
\item[\prince] Princeton University, Princeton, NJ 08544, USA
\item[\riverside] University of Californa, Riverside, CA 92521, USA
\item[\rome] INFN-Sezione di Roma and University of Rome, ``La Sapienza",
     I-00185 Rome, Italy
\item[\salerno] University and INFN, Salerno, I-84100 Salerno, Italy
\item[\ucsd] University of California, San Diego, CA 92093, USA
\item[\sofia] Bulgarian Academy of Sciences, Central Lab.~of 
     Mechatronics and Instrumentation, BU-1113 Sofia, Bulgaria
\item[\korea]  The Center for High Energy Physics, 
     Kyungpook National University, 702-701 Taegu, Republic of Korea
\item[\purdue] Purdue University, West Lafayette, IN 47907, USA
\item[\psinst] Paul Scherrer Institut, PSI, CH-5232 Villigen, Switzerland
\item[\zeuthen] DESY, D-15738 Zeuthen, Germany
\item[\eth] Eidgen\"ossische Technische Hochschule, ETH Z\"urich,
     CH-8093 Z\"urich, Switzerland
\item[\hamburg] University of Hamburg, D-22761 Hamburg, Germany
\item[\taiwan] National Central University, Chung-Li, Taiwan, China
\item[\tsinghua] Department of Physics, National Tsing Hua University,
      Taiwan, China
\item[\S]  Supported by the German Bundesministerium 
        f\"ur Bildung, Wissenschaft, Forschung und Technologie
\item[\ddag] Supported by the Hungarian OTKA fund under contract
numbers T019181, F023259 and T037350.
\item[\P] Also supported by the Hungarian OTKA fund under contract
  number T026178.
\item[$\flat$] Supported also by the Comisi\'on Interministerial de Ciencia y 
        Tecnolog{\'\i}a.
\item[$\sharp$] Also supported by CONICET and Universidad Nacional de La Plata,
        CC 67, 1900 La Plata, Argentina.
\item[$\triangle$] Supported by the National Natural Science
  Foundation of China.
\end{list}
}
\vfill
